\documentclass[runningheads]{llncs}
\usepackage{graphicx}
\usepackage{hyperref}
\usepackage{color,xcolor}
\definecolor{web-link}{RGB}{0,0,139}
\usepackage{enumitem}
\usepackage{stfloats}
\usepackage{caption}
\usepackage{subcaption}
\usepackage{url}

\begin{document}
\title{A 35-Year Longitudinal Analysis of Dermatology Patient Behavior across Economic \& Cultural Manifestations in Tunisia, and the Impact of Digital Tools}
\titlerunning{Impact of Digital Tools on Dermatology Patient Behaviors in Tunisia}
%
\author{Mohamed Akrout\inst{3}\thanks{corresponding author: mohamed@aip.ai} \and Hayet Amdouni\inst{1} \and Amal Feriani\inst{1} \and Monia Kourda\inst{2} \and
Latif Abid\inst{3}}
\authorrunning{M. Akrout et al.}
%
\institute{M.B. Dermatology clinic, Bizerte, Tunisia \and Department of Dermatology, Razi Hospital, Manouba, Tunisia \and
AIPLabs / AIPDerm, Budapest, Hungary}
%
\maketitle              
%
\begin{abstract}

The evolution of behavior of dermatology patients has seen significantly accelerated change over the past decade, driven by surging availability and adoption of digital tools and platforms. Through our longitudinal analysis of this behavior within Tunisia over a 35-year time frame, we identify behavioral patterns across economic and cultural dimensions and how digital tools have impacted those patterns in preceding years. Throughout this work, we highlight the witnessed effects of available digital tools as experienced by patients, and conclude by presenting a vision for how future tools can help address the issues identified across economic and cultural manifestations. Our analysis is further framed around three types of digital tools: ``Dr. Google'', social media, and artificial intelligence (AI) tools, and across three stages of clinical care: pre-visit, in-visit, and post-visit.

\end{abstract}
\section{Introduction}
In this accelerating and expanding digital age, information technologies and image processing techniques have thrown us into an inevitable ``iconosphere'' whose psychological, epistemological, and cultural impacts are undeniable \cite{sierra2015corporeal}. Amongst the various industries where the digital age has transformed the respective processes, products, and services, healthcare has significantly lagged all other industries despite it having the greatest potential, both in terms of the ability to fundamentally improve the delivery of care as well as the overall impact of that to civilization. Despite the connectivity potential and advanced digitization in this new age, patients are still at the mercy of inaccessibility to accurate clinical care in a timely or convenient manner. The potential transformative impact of this digital age is especially exciting, yet thus far disappointing, in lower socio-economic regions of the world, where economic and cultural barriers uniquely inhibit patients' access to the requisite care.

A patient's healthcare ``journey'' has been significantly impacted by software technology \cite{ginige2018transforming}. Patient ``empowerment'' has been augmented due to the easy capture of skin lesions, taken by mobile phones or digital cameras from a distance, and the proliferation of those ``macroscopic'' images in search engines and social media. This empowerment by technological advances is disseminated and encouraged by the media, which is driving patients to be more demanding in their education and in the vocalization of their care with their physicians. 

With healthcare challenges facing the world significantly worsened by COVID, there is an increasing imperative to better analyze and understand how digital tools can transform healthcare across various economic and culturally distinct regions. To effectively design and deploy such tools, it is important that we analyze historical patient behavior across the three clinical stages (pre-, in-, and post-visit) over an extended time frame, and in unique cultural and economic landscapes. This is especially the case in underdeveloped countries such as Tunisia, where healthcare services have more primitive infrastructure and resources. In these regions, digital adoption further lags more developed regions but there is greater potential to improve the current standard of care and address unique economic and cultural barriers \cite{world2013african,azevedo2017state}.

By studying patient behavior over the last 35 years through the consistent lens of two leading dermatologists in Tunisia, we identified that the first 25 of the 35 years mainly experienced slow and minimal change. However, the last 10 years witnessed a significant change due to the introduction of expansive digital tools such as social media, Google search, reverse image searches, and AI tools. Over that time frame, we also witnessed an economic shift in Tunisia after the revolution in 2010-2011, and a significant deterioration in the average household income and level of access and quality of both public and private healthcare.

\noindent This paper makes the following contributions:
\begin{itemize}
    \item We describe the change in patients' behavior over 35 years in Tunisia, segmented into two periods of 1987–2012 and 2012-2022, and we pattern this behavior across coinciding economic and cultural environments.
    \item We discuss and highlight the witnessed effects of digital tools on patient behavior and incorporate that impact across the three stages of clinical care (pre-, in-, and post-visit).
    \item We present a case for how we envision digital tools, such as ``\href{https://semmelweis.aip.ai/en}{\textcolor{web-link}{\texttt{AIPDerm}}}'', can positively address Tunisia's, and similar regions', economic and cultural issues, as well as significantly evolve the current standard of how digital tools impact the patient's behavior across the three clinical stages.
\end{itemize}






\section{35 years of dermatology patient behavior across all three clinic stages}

We analyze patient behavior in a single region of Tunisia across 35 years by examining the direct experiences of two dermatologists who continuously practiced in the region over that period. We segment the observed patient behaviors and perceptions into the three clinical stages: pre-, in-, and post-visit. Within each segment, we analyze the shift of those behaviors and perceptions over the 35-year period, and more specifically, the more notable period of 2012-2022. 

\subsection{Pre-visit Perception}
Preceding the knowledge-at-your-fingertips era of search engines, it was significantly difficult to access any breadth or depth of medical information, constraining patients to almost exclusively trust the diagnosis of the dermatologist. Following the launch and evolution of search engines, patients were more quickly catalyzed at the slightest illness manifestation or concern, with patients now empowered to match their symptoms and be pre-emptively guided on the possible diagnoses, the urgency of the pathology, and the therapeutic possibilities. This empowerment of manually navigating search engines for textual matching of symptoms and conditions has more recently evolved into taking photos of their skin lesions and using reverse image search functionality to find cases with similar visual representations and the corresponding possible skin diagnoses.

Around the 2010-2011 period, Facebook gained significant scale and reach in Tunisia, with this adoption having been validated extensively in the press and public domain given the pivotal role it played in the Tunisian and other ``Arab Spring'' uprisings \cite{allagui2011arab}. On the dermatology side, Tunisian citizens are constantly engaged on Facebook groups to exchange opinions at different engagement levels about $i)$ existing medical prescriptions from prior medical visits, $ii)$ personal judgment about the personality, professionalism, availability, and expertise of the respective dermatologist, and $iii)$ personal experiences with specific treatments for a given pathology (e.g., acne).

While these discussions can represent useful natural language processing datasets for future AI tools and guide the design of clinical trial recruitment, prevention campaigns, and chronic disease management \cite{della2019health}, the quality of shared medical information on social media suffers from the lack of reliability and confidentiality \cite{moorhead2013new}. As the working class's household income and purchasing power have been falling since the 2010-2011 revolution, many patients rely on their discussions on social media to self-educate and self-medicate, exposing patients to various risks such as mistreatment and triggering rashes or drug toxicity. This is in contrast to the pre-digital stronghold era (maturity of search engines and scale-up of social platforms), which had materially lower risk exposure of erroneous and dangerous self-education or self-medication, and where dermatologists were granted more exclusive trust in medical determinations. In order to mitigate some of these risks and limitations with basic self-education, Tunisian website alternatives (e.g., \href{https://med.tn}{\textcolor{web-link}{\texttt{med.tn}}}, \href{https://tobba.tn}{\textcolor{web-link}{\texttt{tobba.tn}}}) offer a dedicated medical forum where subscribed doctors can answer medical questions posted by patients. However, doctors are asked to not provide a specific diagnosis, with their answers instead of revolving around reporting the severity of the patient's case, thus providing a more triaging type of guidance.

Certain skin conditions can be embarrassing or culturally sensitive, such as sexually transmitted diseases (e.g., venereal vegetations, genital herpes, genital candidiasis), which have been, and continue to be, very taboo culturally in Tunisian society. In these cases, most Tunisian patients have notably preferred sending a photo of their lesion accompanied by a textual description of symptoms. This optionality is something that has been enabled by the availability of mobile devices and platforms, and whose absence rendered a significant proportion of the population at the mercy of rigid cultural norms, often resulting in undiagnosed or untreated conditions.
\subsection{In-visit Perception}
Across the clinical time frame of 35 years of experience, we had extensive exposure to both public hospitals and private clinics in Tunisia. We characterize the profile and mindset of dermatology patients during medical visits and classify them into the following (with potential overlapping) categories.\vspace{0.15cm}

\noindent\textbf{Patients with higher education background} \quad This category includes patients with higher-education and scientific backgrounds (e.g., engineers, physicians, architects) or with a background in studying advanced literature (e.g., lawyers, journalists, judges, administrative executives). They expect the dermatologist to direct them towards the right etiological and therapeutic approach after they self-administered and pre-established a survey about their symptoms and possible diagnosis from Google searches and other digital tools. They also keep track of high-resolution images and videos of their skin lesions (e.g., angioedema and urticaria) which simplify and accelerate the diagnostic process of the dermatologist. This category of patients perceives the role of dermatologists as the only single and final source of providing trustworthy diagnoses and treatments. These patient profiles also tend not to over-argue or question the dermatologist's determination of appropriate next steps and treatments (e.g., blood test before drug prescription).\vspace{0.15cm}

\noindent\textbf{Patients with limited income} \quad This category is the one that significantly utilizes social media and Google searches, with those tools most impacting their patient journey through the three clinical stages. These patients try to actively avoid paying medical consultation fees and can excessively opt for self-medication until a friend or pharmacist urges them to consult a dermatologist (e.g., in front of a mole growing in size out of concern of a melanoma diagnosis) or until their condition worsens. For most witnessed patients in this category, dermatologists in Tunisia are seen as the last resort after exhausting all self-medication methods. Over the past decade and with the worsening economic landscape since the 2010-2011 revolution, there has been an increase in patients that fit into this category and who actively seek to avoid the medical fees of multiple visits. We have noticed that they tend to consult a dermatologist for aggregated medical concerns within a single medical visit, which implies that they are allowing for a build-up of illness and manifestations over an extended period of time before acquiescing to paying for fees associated with a single visit. They view this as allowing them to obtain a ``multiple treatment package'' without the intention of treating each of the skin diseases properly and within the appropriate time frame. However, this practice is not prohibited by Tunisian law and tends to drive awkwardness during dermatologist interactions.

\noindent\textbf{Patients who already ``know'' the diagnosis} \quad While it is true that a diagnosis can sometimes seem obvious (e.g., vitiligo, common wart, acne, or burn), patients can often have overly inflated confidence in some of these cases and unnecessarily argue with dermatologists on their medical decisions. The patient consults to ensure possible therapeutic arrangements, their effectiveness, and their aesthetic aspects. Patients often perceive the role of dermatologists as one that is oriented toward frequent follow-ups and very personalized medical advice, without which the patient is not satisfied. This is because the patient can unintentionally not consider the dermatologist's actual core diagnosis as part of the overall medical visit. The role of the dermatologist is to reassure the patient and track them regularly according to the severity of the disease, its location, its extent, and its treatment punctuality. Given that patients can be overconfident about their diagnosis, they tend to ask questions to better understand any discrepancy between the dermatologist's opinion and what they have read on social media or ``Dr. Google''. Questions such as ``we know the diagnosis, so why are you asking for a blood test?'' or ``I saw similar photos to my case on Facebook; why are you complicating my case?'' are not uncommon.

\noindent\textbf{Patients with skin conditions in intimate body locations}\quad Patients in Tunisia having genital skin lesions often refuse to be physically examined in person. They usually come with multiple photos taken before their visit and expect the dermatologist to give a prescription purely based on analyzing the pre-captured photos, as they believe that ``they have nothing else'', and that it must be sufficient for the medical examination. This request goes against the medical code of ethics but is common behavior in Tunisia given the cultural backdrop and sensitivity in the region around intimate body locations and taboo topics such as sexual activity and disease contraction. These embarrassing situations can leave doctors with a dilemma, and often find themselves engaged in a time-consuming cultural discussion in order to convince the patient into accepting the appropriate medical examination. For example, in the case of Behcet's disease, when a patient photographs their genital ulcers and asks the dermatologist for the therapeutic solution without being examined, the dermatologist must explain to the patient that it is imperative to inspect for medical ulcers, pseudo-folliculitis, and even carry out a neurological examination. Predictably, based on the cultural and sexual sensitivities on this front, our discussions with both male and female dermatologists in Tunisia confirm that this scenario is especially frequent when the patient and dermatologist are of opposite genders. In general, we observe that patients in Tunisia actively seek to validate their skin conditions in intimate locations by sending their photos via messaging apps before being obliged to visit the dermatologist in-person. It is worth noting that dermatologists may find themselves in various delicate situations that could give rise to legal interventions, such as the case of a molluscum of the perianal region of a child suggesting sexual abuse. In such medico-legal situations, taking photos becomes imperative and mandatory, but can also be problematic in front of some parents who opposite it in order to avoid cultural shame and stigmas.\vspace{0.1cm}

\begin{figure}[h!]
     \centering
     \hspace{-1.5cm}
     \begin{subfigure}[h]{0.5\textwidth}
         \centering
         \includegraphics[scale=0.035]{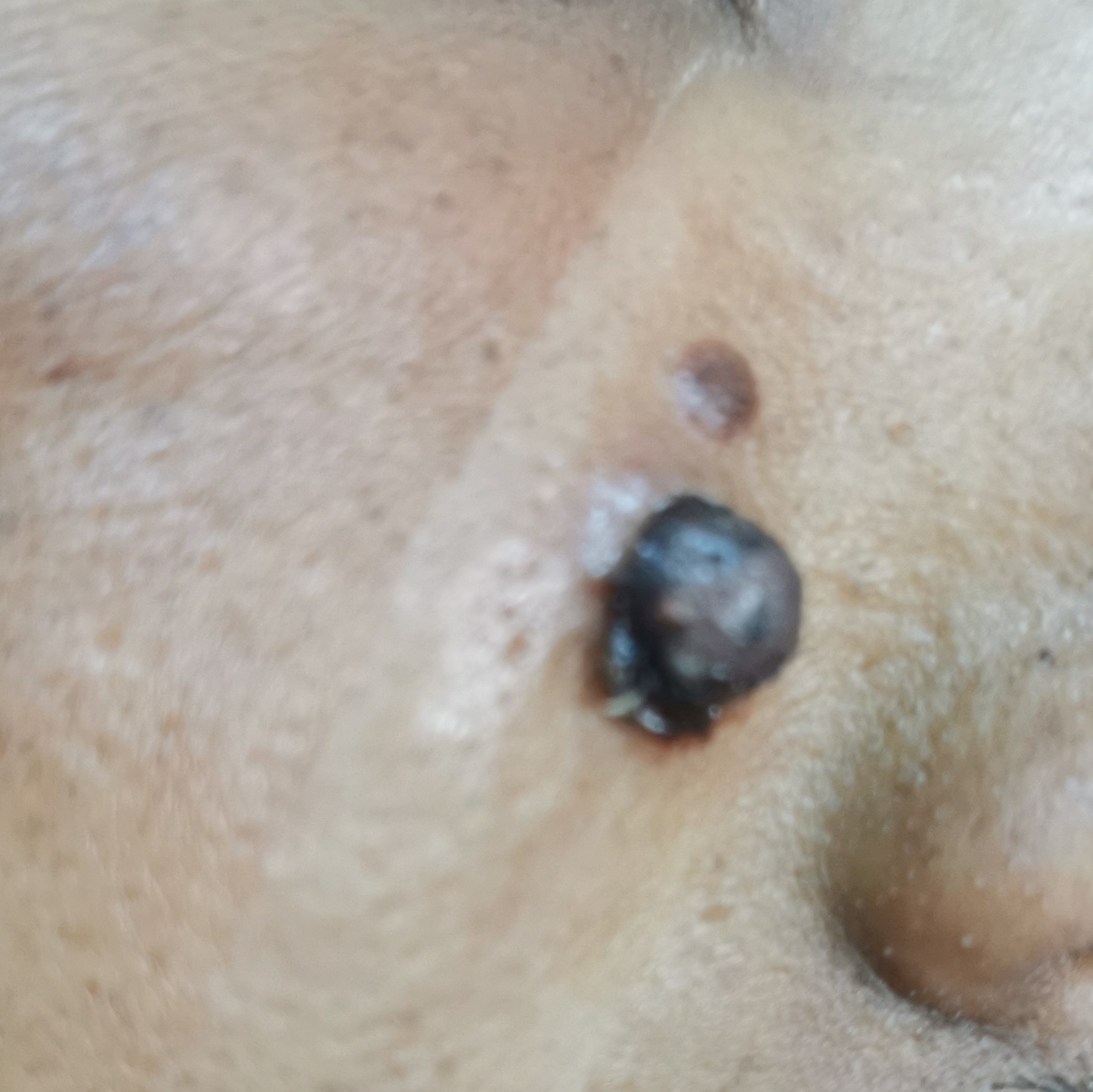}
         \caption{}
         \label{fig:image-before}
     \end{subfigure}
     \begin{subfigure}[h]{0.3\textwidth}
         \centering
         \includegraphics[scale=0.0205]{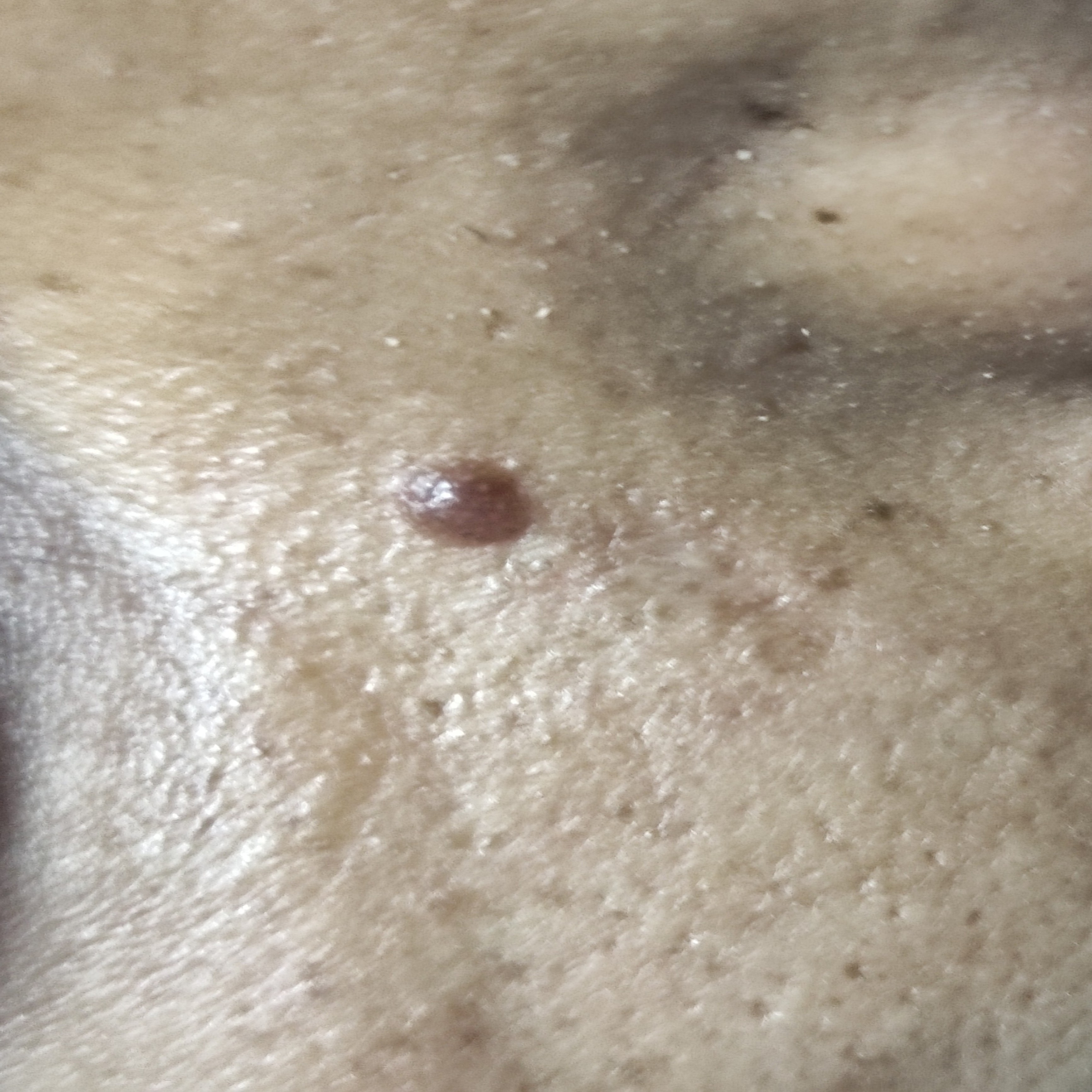}
         \caption{}
         \label{fig:image-after}
     \end{subfigure}
        \caption{An image of a nevus surrounded by a line of a fishing rod in (a) and its evolution after two months of treatment in (b).}
        \label{fig:before-after-example}
        \vspace{-0.7cm}
\end{figure}

\noindent\textbf{Uncooperative patients}\quad Patients with lower intellectual levels, educational backgrounds, or with limited job and career prospects, are often seen spending time on Facebook groups and other social media platforms looking for depigmenting miracle creams for their melasma, and doing so without the more basic concern of worrying about having a sunscreen protecting against ultraviolet type A (UVA), ultraviolet type B (UVB) and blue light. Some of these patients are often seen consulting dermatologists to ``achieve'' their cosmetic dreams that they conjured in their imaginations through sporadic reading of various online sources, and thus have established unreasonably high expectations on what can be achieved and over short periods of treatment \cite{thomas2020usage}. In such cases, many patients are quick to claim that the treatment must not be effective. It has therefore proven prudent for dermatologists in Tunisia to take regular photos showing the temporal evolution of the disease, and present those depictions as clear visually convincing evidence whenever the patient questions the improvement of their skin condition or the inefficacy of the prescribed treatment. Fig. \ref{fig:before-after-example} depicts the case of a patient who visited his hairdresser to shave his beard. The hairdresser suggested to the patient to stop the nevus's volume from increasing by surrounding it with a line of a fishing rod. Fig. \ref{fig:image-before} shows the photo of the patient during his first medical visit where the nevus is surrounded by a line of a fishing rod. The nevus shown in Fig. \ref{fig:image-after} illustrates its evolution after two months of treatment. The patient claimed that the state of his nevus did not improve, yet it was sufficient for the dermatologist to show him his photo in Fig. \ref{fig:image-before} from his first visit to convince him of the improvement.\vspace{-0.4cm}

\subsection{Post-visit Perception}

Patient follow-ups can be almost as important as the actual medical visit, given that they help ensure patients have adhered to their treatments and in tracking the treatments' effects. Given the observed historical tendency for patients to potentially neglect in-person follow-ups, digital tools have shown utility in enabling convenient remote check-ins for post-visit care and tracking. For example, the patient can use digital tools to send images of the progress of their burn to their dermatologist, allowing for remote monitoring as well as treatment adjustments if necessary.
A distinct phenomenon observed in Tunisia is related to the differences between the prescribed treatment by dermatologists (and doctors in general) and the one ultimately provided by pharmacists. These differences can be due to the lack of availability of specific drugs or brands in pharmacies, which is a simple and logical justification. However, the most common reason for the difference in what is prescribed is explained by the unfortunate unfolding competition between pharmacists, who can withhold patients' prescriptions until the unavailable drugs are in-stock again, even though the specifically needed drugs are already available in other pharmacies. In such situations, the ideal patient contacts the dermatologist to verify if the drugs they obtained from the pharmacist are equivalent to the ones prescribed. However, due to the lack of sensitivity and awareness in most cases, the patients will start their treatment without giving adequate attention to the changes in the prescription made by their pharmacists.
Given these more nuanced and technical problems, which are seldom present in more developed countries, dermatologists in Tunisia often find themselves playing the role of ``medical policeman''. For example, the dermatologist can instruct their patient to take a photo of their treatment before beginning to use it to allow for the dermatologist to confirm what has been issued by the pharmacist. A similar issue with cosmetic products that has become more prevalent in recent years is when a patient exchanges drugs with friends, especially when dealing with budget limitations, thereby neglecting the specific prescribed dosage of drugs from the dermatologist for a personalized and optimal therapeutic result.
From this perspective, the perceived role of dermatologists revolves around checking treatments provided by pharmacists and answering any questions, which is often expected for free in cases of complications and uncertainty.

\section{The Historical Evolution of Digital Tools \& Their Patient-Behavior Impact Across Clinical Stages}


The adoption of digital tools has evolved significantly over the past ten years, with distinct progress impacting patient behavior as compared to the 25 years prior to that. However, the full potential of digital tools in the healthcare industry is still far from being realized, with historical implementations largely centering around broader communication and information accessing platforms. While this improves patients' access to any kind of relevant medical information and provides the opportunity to discuss with fellow patients, these tools can be very high in risk of mis-education or mis-treatment, and can hinder dermatologists' efforts in providing optimal care. We investigate ``Dr. Google'', social media, and AI tools, the three major digital tools that were used over the last 10 years in Tunisia, and assess their impact across the clinical stages.\vspace{-0.2cm}

\subsection{Social Media}
The usage of social media in Tunisia, in particular Facebook, has undergone an astonishing growth over the last ten years, and today, Facebook drives 92.65\% of Tunisian social media traffic in the month of June 2022, with similar levels over the last 12 months; Facebook's dominance significantly beats other platforms in that category which includes YouTube, Twitter, Instagram, Pinterest, LinkedIn, Reddit, and other platforms \cite{websiteStat}. Popular applications such as Facebook, Messenger, and Instagram are freely and instantly accessible and provide user-friendly platforms to search and share medical information. Further, Tunisian patients also rely on Facebook groups and pages to seek other patients' opinions and ratings of different dermatologists or clinics. Such platforms offer support systems where disease-focused groups or communities can be created to share experiences and provide psychological support. Facebook and Messenger usage is not restricted to patients, dermatologists also utilize such tools to discuss and solicit each other's opinions and potential diagnoses on more complex or ambiguous cases \cite{masoni2020whatsapp}. Furthermore, social media, and messaging apps facilitate useful patient-doctor and doctor-doctor communications. If the doctor has an online presence, such tools can be used to provide clinic communications, virtual assistance and facilitate follow-ups. Social media platforms have been generally used by Tunisian dermatologists to raise awareness, encourage preventive or misguided behaviors, and offer advice and information on skin diseases to help minimize the risks of self-education and self-treatment. However, the availability of dermatologists to directly regulate content on such platforms is minimal, and their impact is unscalable given the broad usage of those platforms by the general population.

Unfortunately, social media is not always a reliable source of information, especially with complex and highly-personalized diagnoses and treatments, and has been shown to quite easily and quickly disseminate false or misleading information to widespread audiences. Such platforms are often used by commercial organizations or sales-representatives seeking to aggressively promote products for profit, which can lead to harmful misinformation or consequences \cite{smailhodzic2016social}.\vspace{-0.25cm}

\subsection{``Dr. Google''}

Search engine market share in Tunisia is heavily dominated by Google, with 94.95\% of the total search engine market as of June 2022  \cite{websiteStat}. With around 72\% of residents having access to the internet, and with over 90\% of the population having at least one mobile phone, accessibility to the internet and information is quite high, enabling most patients to consult the internet before visiting a dermatologist. Patients usually seek information through searching their symptoms or using more advanced search tools like reverse image searches. The latter is a software tool that enables users to find visually similar images simply by uploading a photograph of the skin lesion. While these tools are used in both developed and underdeveloped countries, dermatologists in Tunisia have emphasized that patients in underdeveloped or economically challenged regions, such as Tunisia, have been driven to rely heavily on these online tools to solely self-medicate, deferring a consult with a specialist to when symptoms significantly deteriorate or misalign with their self-diagnosis.

The ability for patients to seek medical information through various online search engines outside their clinical visit can be beneficial but can also pose significant risks. Self-educating through online resources can facilitate more comprehensive communication between patients and dermatologists, encourage the patient to ask more questions, alert the dermatologist to relevant aspects of their case, and become more engaged with their treatment across the clinical stages. However, the vast amount of information online can be overwhelming, and the ability to ascertain an accurate online diagnosis is low and dangerous, yet is unfortunately still relied upon by patients in deferring visits, especially in less developed or economically challenged countries.


\subsection{The first AI tools}

AI is playing a pivotal role in advancing e-health solutions for dermatology \cite{gomolin2020artificial,rajpara2009systematic,akrout2019improving}. Dermatology is a suitable high-value use case for AI, as it is a visual-dependent specialty and given that numerous image recognition algorithms have already been proven to be effective in detecting skin diseases and cancers \cite{esteva2017dermatologist}. This progress in AI has also fueled the design of several web or mobile image-based diagnosis apps for skin conditions such as
``\href{https://semmelweis.aip.ai/en}{\textcolor{web-link}{\texttt{AIPDerm}}}'', which has had significant success in Europe. These AI-based tools are beneficial for various expertise levels within a clinic, from senior dermatologists to junior nurses, and help distinguish possible skin conditions and provide more sophisticated differential diagnosis possibilities along with ancillary information. AI-based diagnostic tools also play an important role in teledermatology given that they can help patients avoid long waiting periods and pre-identify and triage the severity of their skin condition.

Currently, most dermatologists in Tunisia do not use AI tools for a second opinion. We asked more than 300 dermatologists about their reasons for not yet adopting AI tools, and more than 96\% were worried about the following two aspects of these applications:
\begin{itemize}[leftmargin=*]
    \item \textit{Privacy concerns about users' health data}: most applications indicate they would not use uploaded images to target advertising, and would only save them to further train their classification algorithms, if users gave them explicit permission to do so. However, dermatologists in Tunisia do not have a legal framework established for digital dermatology, and at this early juncture, still feel insecure about adopting these applications.
    \item \textit{Limited coverage of skin diseases}: most dermatology AI applications are limited in scope of capabilities as they are narrowly targeted toward specific skin diseases such as skin cancers. In other words, they do not cover a wide enough diagnostic capability supporting the majority of skin diseases and cancers.
\end{itemize}

\begin{figure}[h!]
\centering
\includegraphics[scale=0.07]{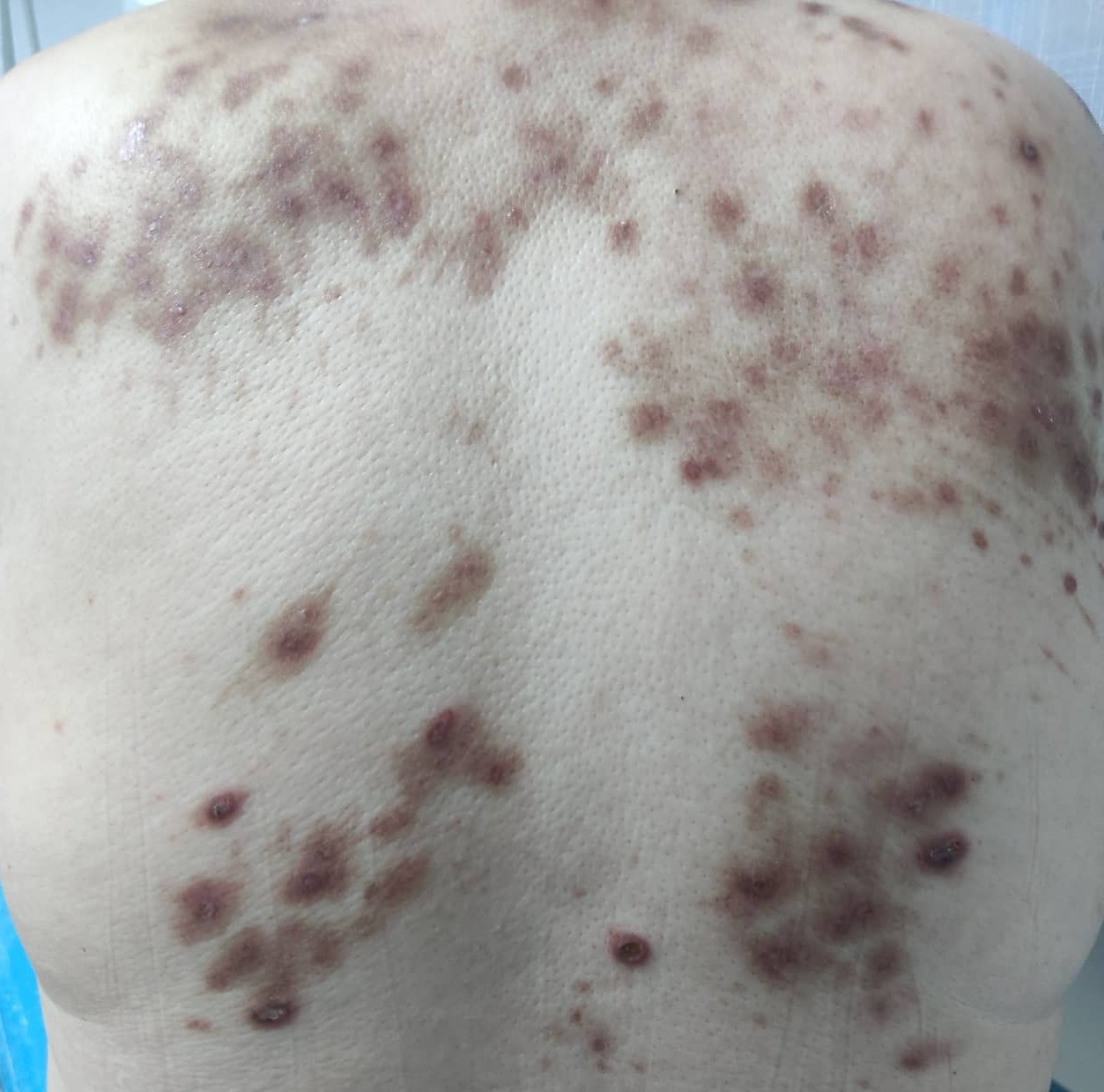}
\caption{The back of a patient with pathomimia disease triggered by a psychological concern.\vspace{-0.8cm}} \label{fig:tough-case-for-AI}
\end{figure}

Remote triaging solutions based on AI tools offer the potential for addressing challenges in many low- and middle-income countries in Africa, or similar economic and culturally inhibiting regions, especially if offered at scale \cite{shuvo2015ehealth}. However, the need for in-person visits remains irreplaceable in many cases, and where the need for precise and nuanced knowledge of medical context is critical. Fig. \ref{fig:tough-case-for-AI} shows one of those cases where the back of the patient has well-defined, painless but itchy, multiple lesions, such as scratches, ulcerations, or cuts. The patient uses a brush to scratch her back. All AI applications we have tested failed this case as it looks like skin picking. This is because the correct diagnosis, pathomimia (a.k.a., factitious dermatosis), is self-induced and often difficult to find if the dermatologist does not consider the psychological profile of the patient, which is confirmed by the opinion of the psychiatrist.\vspace{-0.25cm}

\section{The Future: How Can Advanced Digital Tools Optimize Patient Behavior Across All Clinical Stages and Address Economic \& Cultural Barriers?}

Although we have witnessed transformative evolutions in digital connectivity tools over the last few decades, healthcare remains stagnated in a world of under-adoption and misutilization. The available tools are constantly evolving to be far more sophisticated and tailored than the basic platforms broadly used today. We have begun to see the potential of these future tools in addressing the weakness and limitations of past tools, as global efforts in developing dermatology AI tools have been substantial. 
With respect to the future capabilities of AI systems, we believe its core features should address the patient care and digital tool issues described in this paper. One of the issues is dangerous self-education, and the vast degree of error and risk associated with that. With a system that utilizes sophisticated diagnosis AI modules such as a visual image diagnosis system, the system can minimize the associated error risk by providing a highly accurate diagnostic tool that is not at the mercy of patient diligence. It can be accessed by patients, with results sent to the clinic for oversight, and used as a high-confidence triaging and education tool benefiting both the pre-visit and in-visit stages. For the post-visit stage, a patient management platform can offer a patient portal that provides direct information about their diagnosis and treatments along with detailed reference material for contextual explanations, minimizing the possibility of speculation, skepticism, and non-adherence. This platform can also provide the ability for patients to asynchronously submit questions to the clinic regarding any concerns, minimizing misinformation and ensuring patient actions and understandings are carefully managed by the clinic. This system will also minimize patients’ delays in addressing a possible ailment due to economic reasons and false confidence from self-education, by instantly providing a sophisticated triage and education tool at no incremental overhead costs due to its scalability. The system will also allow dermatologists and their staff to focus their efforts on management and oversight of a single platform while allowing for the ability to follow up with patients on obtaining correct prescriptions and subsequent treatment adherence, as well as enable remote condition monitoring post-visit, which more efficiently qualifies the need for future in-person visits. Such a system has already been comprehensively developed, clinically validated, and successfully deployed at a country-wide level in Europe -- ``\href{https://semmelweis.aip.ai/en}{\textcolor{web-link}{\texttt{AIPDerm}}}', which is based in Hungary and Spain.

Economically, in regions where fiscal stability is sensitive, we believe the scalability and accessibility of such platforms can be transformative. Culturally, we believe such a system is especially relevant in regions such as Tunisia, where longstanding cultural norms and sensitivities can inhibit the requisite treatment for many. In such regions, this can manifest within households where female members can be prohibited from seeing a physician by the male patriarch of the family on the basis of privacy, protection, and cultural sensitivities. We believe that the system can help mitigate such issues by allowing patients to localize the lesion in an image, minimize the inclusion of other body parts, and not require the patient’s physical presence, allowing for remote triaging of their case. Overall, we believe this envisioned system can largely solve the aforementioned patient behavior issues, including cultural and economic constraints, and reduce the perversion of existing basic tools and social media platforms.\vspace{-0.1cm}

\section{Conclusion}
\vspace{-0.1cm}
Patient behavior in dermatology has undergone significant changes over the past 35 years, predominantly over the last decade, driven by significant transformation and adoption of digital tools. Patient behavior is analyzed across three clinical stages: pre-visit, in-visit, and post-visit, and is classified into several profiles that reflect observed patient behaviors. The causes of different patients' behaviors are not singular and this work illuminates the predominantly witnessed patterns in Tunisia. Digital tools have materially impacted how patients behave in all three clinical stages, with various economic and cultural aspects specific to Tunisia also contributing to the patterned patient behavior. While basic digital tools have exposed patients to risks of self-education and self-medication, there is clear potential in minimizing those risks and in mitigating the economic and cultural concerns of Tunisia and various regions globally. The aim of this paper was to describe patient behavior over 35 years in Tunisia across the three clinical stages, while analyzing the coinciding effects of economic and cultural manifestations, and the overarching impact of digital tools. We culminated by detailing how digital tools can mitigate Tunisia’s, and similar regions’, economic and cultural constraints, and how specific innovations can significantly evolve the current standard of how digital tools impact patient behaviors and outcomes.

\vspace{-0.2cm}

%
%
%
%

\bibliographystyle{splncs04}
\bibliography{references.bib}

\end{document}